\documentstyle[12pt]{article}
\begin{document}
\rightline{Dilaton Gravity}
\bigskip
\begin{center}
{\Large\bf Rotating Dilaton Solutions
  in 2+1 Dimensions

}

\end{center}
\hspace{0.4cm}
\begin{center}
{Sharmanthie Fernando \footnote{fernando@nku.edu}}\\
{\small\it Department of Physics}\\
{\small\it Northern Kentucky University}\\
{\small\it Highland Heights}\\
{\small\it Kentucky 41099}\\
{\small\it U.S.A.}\\

\end{center}

\begin{center}
{\bf Abstract}
\end{center}

\hspace{0.7cm}{\small

We report a three parameter
family of
solutions for
dilaton gravity in 2+1 dimensions
with
finite mass and  finite
angular momentum.
These solutions
are obtained by a
compactification
of vacuum solutions in 3+1 dimensions
with cylindrical symmetry.
One class of
solutions  corresponds
to conical singularities and the
other leads to  curvature singularities.
  }

\vspace{1.0cm}


\newpage

\section{Introduction}
Three dimensional gravity has provided us with many
important clues about higher dimensional physics.
String theory, which seems to be the best candidate
available for a consistent theory of quantum gravity,
requires studies of low dimensional
  string effective actions.
In this respect, dilaton gravity in 2+1 dimensions
deserve further attention since it arises from a low
energy string effective theory.

In one of the earlier works, Shiraishi\cite{shiraishi}
found a family of static multi centered solutions for
Einstein-Maxwell-dilaton
gravity. Park and Kim \cite{park}
constructed general static axially symmetric
solutions to the same model
by dimensional reducing to two dimensions.
In recent times, lot of attention has been given
to gravity in 2+1 dimensions
with a negative
cosmological constant due to the existence of
black hole solutions \cite{banados}. Modification
of this black hole
with a dilaton and Maxwell's fields
have lead to many interesting results.
Static charged black holes
and spinning
black holes
in anti-de Sitter space by Chan and Mann\cite{chan2}\cite{chan1},
spinning solutions with self dual electromagnetic fields
by Fernando\cite{fer},
black holes in  generalized dilaton gravity with
a Brans-Dicke type parameter by S\'{a} {\it et.al.}\cite{lemo},
magnetic solutions by Koikawa {\it et.al.}\cite{koi} are
some of the work related to dilaton gravity
in anti-de Sitter space.

In this paper we present an interesting class of
dilaton solutions arising from four
dimensional gravity. This is achieved by a
compactification of
vacuum solutions with
cylindrical symmetry.
These new set of solutions are different from
the ones presented above due its
direct relation to four dimensions.
Our final motivation in studying
three dimensional gravity is to gain
further understanding of higher dimensional
world. Therefore these solutions are important
since it provide us with a
clear understanding of
how the 4D gravity and 3D gravity are related to each other.

We have structured the paper as follows.
In section 2 we will give a brief introduction
to four dimensional vacuum solutions with
cylindrical symmetry and compactify them to obtain dilaton solutions
in three dimensions. In section 3
a prescription to compute the mass and the angular momentum
is highlighted.
In section 4 we will study static
solutions. In section 5 we will study rotating solutions
and finally we will conclude.

\section{Vacuum Solutions with Cylindrical symmetry
in Four dimensions and its Compactification}

Cylindrical symmetrical space-times
in four dimensions consists of  isometries generated
by two commuting space-like Killing vectors.
If the solutions are stationary then the space-time
admit another Killing vector along the time axis.
General stationary cylindrical symmetric
line element with three Killing vectors
$\partial_t, \partial_z,$ and $\partial_{\varphi}$ can be written as,
\begin{equation}
ds^2 = e^{-2U} \left( e^{2K} ( dr^2 + dz^2 ) + r^2 d \phi^2 \right)
-e^{2U} ( dt+ A d \phi)^2
\end{equation}
where $U, K$ and $A$ are functions of $r$ only.
The general stationary vacuum
cylindrical solutions to pure
Einstein action
\begin{equation}
S =    \int d^4x \sqrt{-G} R
\end{equation}
are  given by,
$$e^{2U} = r(a_1r^n + a_2 r^{-n})$$
$$e^{2K-2U} = r^{\frac{n^2-1}{2}}$$
$$A =  \frac{c}{na_2}  \frac{r^n} { (a_1r^n + a_2 r^{-n})} + b$$
\begin{equation}
c^2 =- n^2 a_1 a_2
\end{equation}
The complex constants, $n, c, a_1, a_2, b$
have to be chosen such that the metric is real \cite{kramer}.
In this paper, we will choose all these constants to be real.

The purpose of this paper is to dimensionally reduce
four dimensional Einstein gravity
to three dimensions
to obtain dilaton gravity. The field content of
the reduced theory would
be gravity, dilaton $\phi$
and the gauge
field $A_{\mu}$. How ever, if we pick the cylindrical
solutions  given above
and treat the $z$ coordinate to
be compact
for the purposes of
equations of motion and their symmetries,
the reduction will yield a theory
in three dimensions with only a dilaton field
coupled to gravity.
To perform compactification along the
$z$ direction, let us
rewrite the above metric in four dimensions as
follows:
\begin{equation}
  ds^2_{3+1} = G_{\mu \nu} dx^{\mu} dx^{\nu}=
  {g}_{ab} dx^a dx^b + e^{-4\phi}dz^2
  \end{equation}
  with
  \begin{equation}
  e^{-4 \phi} = G_{zz} = r^{\frac{(n^2-1)}{2}}
  \end{equation}
  Here $ (\mu, \nu = 0,1,2,3)$
  and $(a,b = 0,1,2 )$ are four and
  three dimensional indices respectively.
  The  dimensionally reduced action in three dimensions
is given by,
  \begin{equation}
  S_{string} = \int d^3x  \sqrt{-g^S} e^{-2 \phi} R^S
  \end{equation}
  Here, $g^S$ corresponds to $g_{ab}$ of the four dimensional
  metric and the metric can be treated to be in the ``string frame''.
  One can perform a conformal transformation
  to bring the metric to Einstein frame as follows,
  \begin{equation}
  e^{-4 \phi}g^S_{\mu \nu} = g^E_{\mu \nu}
  \end{equation}
  This transformation  will lead to the following action and the
  corresponding field equations,
  \begin{equation}
  S_{Einstein} = \int d^3x \sqrt{-g^E} ( R^E - 8 \nabla \phi_{\mu} 
\nabla^{\mu} \phi)
  \end{equation}
  \begin{equation}
R_{\mu \nu} = 8\nabla_{\mu} \phi \nabla_{\nu} \phi
\end{equation}
\begin{equation}
\nabla^{\mu}\nabla_{\mu} \phi = 0
\end{equation}
Hence by starting from the metric given in eq.(1),
one can obtain solutions to three dimensional space-time
as,
\begin{equation}
ds_{Ein}^2 = e^{-2U - 4 \phi} \left( e^{2K} dr^2  + r^2 d \varphi^2 \right)
-e^{2U -4 \phi} ( dt+ A d \varphi)^2
\end{equation}
with  a dilaton given by,
\begin{equation}
\phi = \frac{(-n^2 + 1)}{8} ln (r)
\end{equation}
The above metric can be rewritten in the following form,
\begin{equation}
ds^2 = g_{tt}dt^2 +   g_{\varphi \varphi}d \varphi^2
+ 2g_{\varphi t} d \varphi dt
+ g_{rr} dr^2
\end{equation}
with
$$
g_{rr} = r^{n^2-1}
$$
$$
g_{tt} = - a_1r^{\frac{(n +1)^{2}}{2}} - a_2r^{\frac{(n -1)^{2}}{2}}
$$
$$
g_{\varphi \varphi} = \left( \frac{1}{a_2} - 2b \sqrt{\frac{-a_1}{a_2}} -
b^2 a_1 \right) r^{\frac{(n+1)^2}{2}}
- b^2 a_2 r^{\frac{(n-1)^2}{2}}$$
\begin{equation}
g_{\varphi t} = \left( - ba_2  \right)
r^{\frac{(1-n)^2}{2}}
+ \left( - ba_1 - \sqrt{-a_1/ a_2} \right) r^{\frac{(1+n)^2}{2}}
\end{equation}
This metric and the dilaton $\phi$
satisfy the field equations of eqs.(9) and (10).
In the discussion of the solutions we will pick
$a_2 > 0, a_1=0$ and $n \geq 1$.

\section{Mass and  Angular Momentum
of  the Solutions}
The mass and the angular momentum
of the source of this solution is computed
by following the prescription of Brown and York \cite{bro}
which is briefly described as follows.
If the metric in 2+1 dimensions
is written in the following form,
\begin{equation}
ds^2 = -N^{t2} dt^2 +   \frac{dR^2}{f^2} +
R^2 ( d \varphi + N^{\varphi} dt)^2
\end{equation}
the quasi-local mass $m(R)$
and quasi-local angular momentum $j(R)$
are given by,
\begin{equation}
m(R) = 2 N^t(R) ( f_0(R) - f(R)) - j(R)N^{\varphi}(R)
\end{equation}
\begin{equation}
j(R) = \frac{f(R) \frac{dN^{\varphi}}{dR} R^3} {N^t(R)}
\end{equation}
Here, $f_0$ comes from a background metric which
corresponds to the solution with zero mass.
If $R$ is a function of $r$, as it is in
the case
of  the solutions
described in this paper, $f(r)$ and
$g_{rr}$ are related to each other as
$f(r) =  \sqrt{1/g_{rr}} \frac{dR}{dr}$.
The mass and the
angular momentum
are computed as $R \rightarrow \infty$,
\begin{equation}
M = \lim_{R \to \infty} m(R);   \hspace{1.0cm} J = \lim_{R \to \infty} j(R)
\end{equation}

\section{Static Dilaton Solutions}
In this section we will consider the static
dilaton solutions corresponding to  $b=0.$
Then the  metric simplifies to,
\begin{equation}
ds^2 = - a_2 r^{\frac{(-1+n)^2}{2}}dt^2 +   \frac{dr^2}{r^{(1-n^2)}}
+ \frac{r^{\frac{{(1+n)}^2}{2}}}{a_2}
  d \varphi^2
\end{equation}
\begin{equation}
\phi = \frac{(-n^2 + 1)}{8} ln (r)
\end{equation}

\subsection{ Static solutions for  $n=1$}
For $n =1$, the dilaton vanishes leaving
the following metric.
\begin{equation}
ds^2 = -a_2dt^2 + dR^2 a_2 + R^2 d\varphi^2
\end{equation}
Note that a coordinate transformation
$r= R\sqrt{a_2}$ has been performed.
This is the well known
metric of a point source in 2+1 dimensions
which leads to a conical singularity
at the origin \cite{des} \cite{gott}. Let  us
describe how the conical singularity arises
as follows:
By redefining
$t, R$ and $\varphi$
as
$t' = \sqrt{a_2}t, R' = \sqrt{a_2}R,  \varphi' = \varphi/ \sqrt{a_2}$,
the metric in eq.(21) becomes,
\begin{equation}
ds^2 = -dt'^2 + dR'^2  + R'^2 d\varphi'^2
\end{equation}
Note that the  former periodic coordinate $\varphi$
has the range
$ 0 \leq \varphi \leq 2 \pi$ and the new period coordinate
has the range
$ 0 \leq \varphi' \leq 2 \pi a_2^{-1/2}$.
Hence there is a deficit angle $D$ at the origin
due to the presence of a massive source
given by $D = 2\pi ( 1 - a_2 ^ {-1/2})$
leading to the conical space-time.
Now, to relate the parameter $a_2$ appearing in the
metric to the mass of
the source we will use the prescription given in the section
(3).
The zero mass  metric is chosen to be
the Minkowski space which corresponds to $a_2 =1$,
leading to $f_0 = 1$. In the presence of the source,
$f(R) = a_2^{-1/2}$ and  $N^t(R) = 1$ from eq.(21).
By the definition
in eq.(16),
\begin{equation}
m(R) = M = 2( 1 - a_2^{-1/2}).
\end{equation}
Hence  the mass of the source $M$ and $a_2$ are related by
$a_2 = \left(1 -\frac{M}{2} \right)^{-2}$

\subsection{Static  solutions for $n > 1$}
For $n > 1$, the dilaton has
a non zero value and the metric can be written completely
in terms of $R$ as follows,
\begin{equation}
ds^2 = R^{2(\frac{n-1}{n+1})^2} \left(
-a_2^{\frac{2(n^2+1)}{(n+1)^2}}dt^2 +
16a_2^{\frac{2(n^2+1)}{(n+1)^2}} (1+n)^{-4}
dR^2 \right) + R^2 d\varphi^2
\end{equation}
By scaling  the time as $a_2^{\frac{(n^2+1)}{(1+n)^2}} t = t'$,
the metric simplifies to,
\begin{equation}
ds^2 = -A(R)^2 dt'^2 + \frac{dR^2}{ (A(R)^{-1}a_n)^2} + R^2 d\varphi^2
\end{equation}
where
\begin{equation}
a_n = a_2^{\frac{-(n^2+1)}{(n+1)^2}} \frac{(1+n)^2}{4}; \hspace{1.0cm}
A(R) = R^{(\frac{n-1}{n+1})^2}
\end{equation}
The functions $f(R)$ and $N^t(R)$ can be read from the metric
as,
\begin{equation}
f(R) = A(R)^{-1}a_n ; \hspace{1.0cm} N^t(R) = A(R)
\end{equation}
To compute the mass,  let the reference metric corresponds
to $a_2 =1$ which leads to
\begin{equation}
f_0 =  A(R)^{-1}\frac{(1+n)^2}{4}
\end{equation}
With above preliminaries,
one can compute the mass and the angular momentum to be,
\begin{equation}
M_n =\frac{(n+1)^2}{2} \left( 1 - a_2^{\frac{-(1+n^2)}{(1+n)^2}} \right);
\hspace{1.0cm} J_n = 0
\end{equation}
When $n \rightarrow 1$, the mass $M_n \rightarrow M$ as expected.

The Ricci scalar $Rs$ and Kretschmann scalar
$Kr$ are computed for the above metric as follows:
\begin{equation}
Rs = R_{\mu \nu}g^{\mu \nu} = \frac{(-n^2 + 1)}{8r^{(1+n^2)}}; \hspace{1.0cm}
Kr = R_{\mu \nu \rho \gamma}R^{\mu \nu \rho \gamma}
= \frac{3(-n^2 + 1)}{64r^{2(1+n^2)}}
\end{equation}
For $n > 1$ these scalars have singularities at $r=0$
leading to  a curvature  singularity.  Since
there are no horizons, it is a  naked singularity.
For $n =1$
the scalar curvature is zero
everywhere
leading to the conical singularity at the origin
as discussed in the previous section.
Hence
the conical singularity has turned into a curvature singularity
due to the presence of the dilaton field.
Since the scalar curvature $R \rightarrow 0$ at large $R$,
the space-time is asymptotically flat.

\section{Rotating Dilaton Solutions}
In this section we will consider the solutions with the parameter
$b \not= 0$.  Such solutions can be written as,
\begin{equation}
ds^2 = -\frac{r^{(n^2+1)}}{R^2}dt^2 + \frac{dr^2}{r^{(-n^2+1)}}
+ R^2 \left( d \varphi - \frac{a_2br^{\frac{(n-1)^2}{2}}}{R^2} dt  \right)^2
\end{equation}
with
\begin{equation}
R^2 = \frac{r^{\frac{(n+1)^2}{2}}}{a_2} - b^2a_2r^{\frac{(n-1)^2}{2}}
\end{equation}

\subsection{ Rotating flat solutions with $n =1$}
For $n=1$, the dilaton field vanishes and the solutions
correspond to the following metric.
\begin{equation}
ds^2 = -\frac{r^2}{R^2} dt^2 + dr^2 +  R^2 ( d \varphi -
\frac{a_2b}{R^2} dt)^2
\end{equation}
where $R^2 = ( \frac{r^2}{a_2} - b^2a_2)$.
The conical singularity at $r=0$
still exists. Also
$g_{\varphi \varphi}$
component of the metric becomes
negative
for $r < ba_2$
leading to closed time-like curves
as described in Deser {\it et.al.}\cite{des}
The space-time is flat everywhere since the
scalar curvature vanishes similar to the static flat case.

To compute the mass, the metric can be rewritten completely
in terms of $R$ by a coordinate transformation
$ r^2 = (R^2 + a_2 b^2)a_2$ and $t' =\sqrt{a_2}t$.
\begin{equation}
ds^2 = -\frac{(R^2 + a_2 b^2)}{R^2} dt'^2 + \frac{R^2a_2}
{(R^2 + a_2 b^2)}dR^2 +  R^2 ( d \varphi -
\frac{\sqrt{a_2}b}{R^2} dt')^2
\end{equation}
Considering the Minkowski space-time
as the reference one can compute the mass and the angular momentum
as
\begin{equation}
M = 1 - a_2^{-1/2}; \hspace{1.0cm} J = 2b.
\end{equation}

\subsection{Rotating Dilaton Solutions with $n > 1$}
For $n > 1$, a non zero dilaton field exists which
modifies the flat space-time considerably.
These solutions do have the same scalar invariants
as computed for the static case in eq.(31)
which signals curvature
singularities at $r=0$.
Furthermore  the metric function $g_{\varphi \varphi}$
becomes negative  for $ r < (ba_2)^{1/n}$
leading to closed time like curves.

To calculate the mass
and the angular momentum, one has to rewrite $r$ as a function of $R$.
Due to the nature
of the expression in eq.(32) it is not possible
to find an exact expression for $r$.
Hence,
by a binomial
expansion of eq.(32)
around
$R \rightarrow \infty$
followed by an inversion of the  series lead to,
\begin{equation}
\frac{1}{r} =  \left(\frac{1}{R^2a_2} \right)^{\frac{2}{(1+n)^2}}
- \frac{2a_2^2b^2}
{(1+n)^2} \left(\frac{1}{R^2a_2} \right)^{\frac{2(2n+1)}{(1+n)^2}}
+ .....
\end{equation}
Substitution of $r$ into the functions $N^t(r), f(r), N^{\varphi}(r)$
in eq.(31)
to compute the quasi-local mass and angular momentum $m(R)$,
$j(R)$
and  taking the limit $R \rightarrow \infty$ leads to
following quantities,
\begin{equation}
M =\frac{(n+1)^2}{2} \left( 1 - a_2^{\frac{-(1+n^2)}{(1+n)^2}} \right);
\hspace{1.0cm}  J = 2nb
\end{equation}
Note that in computing the mass we have chosen the reference metric
as the one with $a_2 = 1, b=0$ and $n \not= 1$
which gives $f_0 =  R^{-(\frac{n-1}{n+1})^2}\frac{(1+n)^2}{4}$
as in the static case.

\section{Conclusions}
We have discussed
the properties of  a new three parameter
family of solutions to Einstein-dilaton
gravity in 2+1 dimensions
obtained by a compactification of
stationary cylindrical
symmetrical solutions
in 3+1 dimensions.
  The mass
and the angular momentum of each solution are computed in terms of
the parameters of the solution
$a_2, n$ and $b$.
For $n=1$, the compactified solutions
lead to the well known conical  space-time with
a mass deficit and
time helical structure presented by Deser {\it et.al.}\cite{des}
arising in pure Einstein gravity in 2+1 dimensions.
For $n > 1$,
a non zero dilaton field exists
and the resulting
space-time has curvature singularities. The rotating solutions also
has  closed time like curves.

It is natural to extend this work to
compactify charged cylindrical solutions in four dimensions
to see the relation to
the existing charged solutions in 2+1 dimensional
dilaton gravity. In this context, there
are two solutions which would be interesting to study.
One is the
static cylindrical solutions by Safko \cite{saftko}
and the other is the
solutions of a charged line-mass by Muckherji \cite{muck}.

It would be also interesting  to embed the solutions discussed here
in a supergravity theory arising from a low energy string theory
along the lines of supersymmetric solutions to
three dimensional heterotic string action considered
by Bakas {\it et.al.}\cite{bakas}.
We hope to address these
issues in the future.\\
{\bf Note added in proof:} The authors were informed of related work 
mentioned in references  \cite{lemos1}, \cite{lemos2}, \cite{lemos3}, 
\cite{clem}, \cite{vir}.


\begin{thebibliography}{99}
\bibitem{shiraishi} K. Shiraishi, J. of Math. Phys. {\bf 34} (1993) 1480
\bibitem{park}D. Park and J.K. Kim, J. Math. Phys. {\bf 38} (1997) 2616.
\bibitem{banados} M. Ba\~{n}ados, C. Teitelboim and J. Zanelli,
  Phys. Rev. Lett. {\bf 69} (1992) 1849; M. Ba\~{n}ados, M.
Henneaux, C. Teitelboim and J.
Zanelli, Phys. Rev. D {\bf 48} (1993) 1506.
\bibitem{chan2}K.C.K. Chan and R.B. Mann ,  Phys. Rev.  {\bf D50} (1994) 6385.
\bibitem{chan1}K.C.K. Chan and R.B. Mann ,  Phys. Lett. {\bf B371} (1996) 199.
\bibitem{fer} S. Fernando, Phys. Lett. {\bf B468 }  (1999) 201.
\bibitem{lemo} P.M. S\'{a}, A. Kleber and J.P.S. Lemos, Class. Quan. 
Grav. {\bf 13} (1996) 199.
\bibitem{koi}T. Koikawa, T. Maki and A. Nakamula, Phys. Lett. {\bf B414} (1997)
45.
\bibitem{kramer} D. Kramer, H. Stephani, E. Herlt, M. MacCallum and 
E. Schmutzer,
``Exact Solutions of Einstein's Field Equations'', Cambridge Press (1980)
\bibitem{bro} J.D. Brown and J.W. York,  Phys. Rev.{\bf D47} (1993) 1407.
\bibitem{des} S. Deser, R. Jackiw and G. 'T Hooft, Ann. Phys.
{\bf 152} (1984) 220.
\bibitem{gott} J.R. Gott and M. Alpert, Gen. Rela. \& Grav. {\bf 16} 
(1984) 243.
.
\bibitem{muck} B. C. Muckherji, Bull. Calc. Math. Soc. {\bf 30}, (1938) 95.
\bibitem{saftko} J. L. Safko, Phys. Rev. {\bf D16} (1977) 1678.
\bibitem{bakas} I. Bakas, M. Bourdeau, and G. L. Cardoso, Nucl. Phys. 
{\bf B510}
(1998) 103.
\bibitem{lemos1} Oscar J.C. Dias and Jose' P.S. Lemos, JHEP 01 (2002) 006
\bibitem{lemos2} Oscar J.C. Dias and Jose' P.S. Lemos, Phys.Rev {\bf 
D64} (2001) 064001
\bibitem{lemos3} Oscar J.C. Dias and Jose' P.S. Lemos, hep-th/0110202
\bibitem{clem} G. Clement and A. Fabbri, Class. Quant. Grav. 16 (1999) 323
\bibitem{vir} K. S. Virbhadra, gr-qc/9408035





\end{thebibliography}
\end{document}